\documentclass{amsart}
\usepackage{amssymb}
\usepackage{amsmath}

\theoremstyle{plain}
\newtheorem{theorem}{Theorem}[section]
\newtheorem{lemma}[theorem]{Lemma}

\theoremstyle{definition}
\newtheorem{definition}[theorem]{Definition}

\theoremstyle{remark}

\numberwithin{equation}{section}

\input{tcilatex}

\begin{document}
\title[GELFAND-LEVITAN-MARCHENKO TYPE EQUATIONS]{THE STRUCTURE OF
GELFAND-LEVITAN-MARCHENKO TYPE EQUATIONS FOR DELSARTE TRANSMUTATION
OPERATORS OF LINEAR MULTI-DIMENSIONAL DIFFERENTIAL OPERATORS AND OPERATOR
PENCILS. PART 1.}
\author{J. Golenia}
\address{The AGH University of Science and Technology, Department of Applied
Mathematics, Krakow 30059 Poland}
\email{jnapora@wms.mat.agh.edu.pl}
\author{A.K. Prykarpatsky}
\address{The AGH University of Science and Technology, Department of Applied
Mathematics, Krakow 30059 Poland, and Dept. of \ Nonlinear Mathematical
Analysis at IAPMM, NAS of Ukraine, Lviv 79601 Ukraina}
\email{prykanat@cybergal.com, pryk.anat@ua.fm}
\author{Y.A. Prykarpatsky}
\address{Intitute of Mathematics at the NAS, Kiev 01601, Ukraine, and the
AGH University of Science and Technology, Department of Applied Mathematics,
Krakow 30059 Poland}
\curraddr{Brookhaven Nat. Lab., CDIC, Upton, NY, 11973 USA}
\email{yarpry@bnl.gov}
\date{}

\begin{abstract}
An analog of Gelfand-Levitan-Marchenko integral equations for multi-
dimensional Delsarte transmutation operators is constructed by means of
studying their differential-geometric structure based on the classical
Lagrange identity for a formally conjugated pair of differential operators.
An extension of the method for the case of affine pencils of differential
operators is suggested.
\end{abstract}

\maketitle



\setcounter{equation}{0}

\section{\protect\bigskip Introduction}

Consider the Hilbert space ${\mathcal{H}}=L_{2}(\mathbb{R}^{m};\mathbb{C}%
^{N}),$ $m,N\in \mathbb{Z}_{+},$ and the correspondingly conjugated pair ${%
\mathcal{H}}^{\ast }\times {\mathcal{H}}$ \ on which one can define the
natural scalar product%
\begin{equation}
(\varphi ,\psi )=\int_{\mathbb{R}^{m}}dx<\varphi ,\psi >:=\int_{\mathbb{R}%
^{m}}dx\bar{\varphi}^{\intercal }(x)\psi (x),  \label{1.1}
\end{equation}%
where $(\varphi ,\psi )\in {\mathcal{H}}^{\ast }\times {\mathcal{H}},$ the
sign $"-"$ means the complex conjugation and the sign $"\intercal "$ means
the standard matrix transposition. Take also two linear densely defined
differential operators $L$ and $\tilde{L}:{}{\mathcal{H}}\rightarrow {%
\mathcal{H}}$ ${}$ and some two closed functional subspaces ${\mathcal{H}}%
_{0}$ and $\tilde{{\mathcal{H}}_{0}}\subset \mathcal{H}_{-},$ where $%
\mathcal{H}_{-}$ is the negative Hilbert space from a Gelfand triple
\begin{equation}
\mathcal{H}_{+}\subset \mathcal{H}\subset \mathcal{H}_{-}  \label{1.1a}
\end{equation}%
of the correspondingly Hilbert-Schmidt rigged \cite{Be,BS} Hilbert space $%
\mathcal{H}.$ We will use further the following definition.

\begin{definition}
(see J. Delsarte and J. Lions \cite{De,DL} ). A linear invertible operator $%
\mathbf{\Omega },$ defined on $\mathcal{H}$ \ and acting from ${\mathcal{H}}%
_{0}$ onto $\tilde{\mathcal{H}}_{0},$ is called a Delsarte transmutation
operator for a pair of linear differential operators $L$ and $\tilde{L}:{%
\mathcal{H}}\longrightarrow {\mathcal{H}},$ if the following two conditions
hold:

$\bullet $ the operator $\mathbf{\Omega }$ \ and its inverse $\mathbf{\Omega
}^{-1}$ are continuous in $\mathcal{H};$

$\bullet $ the operator identity
\begin{equation}
\tilde{L}\mathbf{\Omega }=\mathbf{\Omega }L  \label{1.2}
\end{equation}

is satisfied.
\end{definition}

Such transmutation operators were for the first time introduced in \cite%
{De,DL} for the case of one-dimensional differential operators. In
particular, for the Sturm-Liouville and Dirac operators the
complete structure of the corresponding Delsarte transmutation
operators was described in \cite{LS,Ma}, where also the extensive
applications to spectral theory were done. As was shown in
\cite{LS,Ni,Ma}, for the case of one-dimensional differential
operators an important role in theory of Delsarte transmutation
operators is played by special integral Gelfand-Levitan-Marchenko
(GLM) equations \cite{Na,Ma,LS}, whose solutions are exactly
kernels of the corresponding Delsarte transmutation operators.
Some results for two-dimensional Dirac and Laplace type operators,
were also obtained in \cite{Fa,Ni} .

In the present work, based on results of \cite{PSP,PSS,GPSP,SP}, we shall
construct for a pair of multi-dimensional differential operators acting in a
Hilbert space ${\mathcal{H}}$ a special pair of \ conjugated Delsarte
transmutation operators $\mathbf{\Omega }_{+}$ and $\mathbf{\Omega }_{-}$ in
$\mathcal{H}$ and a pair $\mathbf{\Omega }_{+}^{\circledast }$ and $\mathbf{%
\Omega }_{-}^{\circledast }$ \ in $\mathcal{H}^{\ast }$ parametrized by two
pairs of closed subspaces ${\mathcal{H}}_{0},$ $\tilde{{\mathcal{H}}_{0}}%
\subset \mathcal{H}_{-}$ and ${\mathcal{H}}_{0}^{\ast },$ ${\mathcal{\tilde{H%
}}}_{0}^{\ast }\subset \mathcal{H}_{-}^{\ast },$ such that the operators $%
\mathbf{\Phi }:=\mathbf{\Omega }_{+}^{-1}\mathbf{\Omega }_{-}-\mathbf{1}$
from $\mathcal{H}$ to $\mathcal{H}$ \ and $\mathbf{\Phi }^{\circledast }:=%
\mathbf{\Omega }_{+}^{\circledast ,-1}\mathbf{\Omega }_{-}^{\circledast }-%
\mathbf{1}$ from $\mathcal{H}^{\ast }$ to $\mathcal{H}^{\ast }$ \ \ are ones
of Hilbert-Shmidt type, thereby determining via the equalities
\begin{equation}
\mathbf{\Omega }_{+}(1+\mathbf{\Phi })=\mathbf{\Omega }_{-},\text{ \ }%
\mathbf{\Omega }_{+}^{\circledast }(1+\mathbf{\Phi }^{\circledast })=\mathbf{%
\Omega }_{-}^{\circledast }  \label{1.3}
\end{equation}%
the corresponding analogs of GLM-equations, taking into account that
supports of both kernels of integral operators $\mathbf{\Omega }_{+},\mathbf{%
\Omega }_{-}$ and $\mathbf{\Omega }_{+}^{\circledast }$ $,\mathbf{\Omega }%
_{-}^{\circledast }$ are correspondingly disjoint. Moreover, the following
important expressions
\begin{eqnarray}
\mathbf{\Omega }_{+}L\mathbf{\Omega }_{+}^{-1} &=&\tilde{L}=\mathbf{\Omega }%
_{-}L\mathbf{\Omega }_{-}^{-1},\text{ \ }  \label{1.4} \\
(\mathbf{1}+\mathbf{\Phi })L &=&L(\mathbf{1}+\mathbf{\Phi }),\text{ \ }(%
\mathbf{1}+\mathbf{\Phi }^{\circledast })L^{\ast }=L^{\ast }(\mathbf{1}+%
\mathbf{\Phi }^{\circledast })  \notag
\end{eqnarray}%
hold. As in the classical case \cite{Ma,LS,Na}, the solutions to this
GLM-equation also give rise to kernels of the corresponding Delsarte
transmutation operators $\mathbf{\Omega }_{\pm }$ in $\mathcal{H},$ that are
very important \cite{BS,No} for diverse applications.

Another trend of this work is related with a similar problem of constructing
Delsarte transmutation operators and corresponding integral GLM-equations
for affine pencils of linear multi-dimensional differential operators in $%
\mathcal{H},$ having important applications, in particular, for the inverse
spectral problem and feedback control theory \cite{DS}.

\section{Generalized Lagrangian identity, its differential-geometric
structure and Delsarte transmutation operators}

\setcounter{equation}{0}Consider a multi-dimensional differential operator $%
L:{\mathcal{H}}\longrightarrow {\mathcal{H}}$ of order $n(L)\in \mathbb{Z}%
_{+}:$
\begin{equation}
L(x;\partial ):=\sum_{|\alpha |=0}^{n(L)}a_{\alpha }(x)\frac{\partial
^{|\alpha |}}{\partial x^{\alpha }},  \label{2.1}
\end{equation}%
defined on a dense subspace $D(L)\subset \mathcal{H},$ where, as usually,
one assumes that coefficients $a_{\alpha }\in {\mathcal{S}}({\mathbb{R}}%
^{m};End{\mathbb{C}}^{N}),$ $\alpha \in \mathbb{Z}_{+}^{m}$ is a
multi-index, $|\alpha |=\overline{0,n(L)},$ and $x\in {\mathbb{R}}^{m}.$ The
formally conjugated to (\ref{2.1}) operator $L^{\ast }:{\mathcal{H}}^{\ast
}\longrightarrow {\mathcal{H}}^{\ast }$ is of the form
\begin{equation}
L^{\ast }(x;\partial ):=\sum_{|\alpha |=0}^{n(L)}(-1)^{|\alpha |}(\frac{%
\partial ^{|\alpha |}}{\partial x^{\alpha }}\cdot \bar{a}_{\alpha }(x)),
\label{2.2}
\end{equation}
$x\in {\mathbb{R}}^{m}$ and the dot $"\cdot "$ above means the usual
composition of operators.

Subject to the standard semilinear form $<\cdot ,\cdot >$ on $\mathbb{C}%
^{N}\times \mathbb{C}^{N}$ one can write down easily the following
generalized Lagrangian identity:
\begin{equation}
<L^{\ast }\varphi ,\psi >-<\varphi ,L\psi >=\sum_{i=1}^{m}(-1)^{i+1}\frac{%
\partial }{\partial x_{i}}Z_{i}[\varphi ,\psi ],  \label{2.3}
\end{equation}%
where for any pair $(\varphi ,\psi )\in {\mathcal{H}}^{\ast }\times {%
\mathcal{H}}$ expressions $Z_{i}[\varphi ,\psi ],$ $i=\overline{1,m},$ being
semi-linear on ${\mathcal{H}}^{\ast }\times {\mathcal{H}}.$ Having now
multiplied (\ref{2.3}) by the oriented Lebesgue measure $dx:=\underset{j=%
\overrightarrow{1,m}}{\wedge }dx_{j},$ we easily get that
\begin{equation}
\lbrack <L^{\ast }\varphi ,\psi >-<\varphi ,L\psi >]dx=dZ^{(m-1)}[\varphi
,\psi ],  \label{2.4}
\end{equation}%
where
\begin{equation}
Z^{(m-1)}[\varphi ,\psi ]:=\sum_{i=1}^{m}dx_{1}\wedge dx_{2}\wedge ...\wedge
dx_{i-1}\wedge Z_{i}[\varphi ,\psi ]dx_{i+1}\wedge ...\wedge dx_{m}
\label{2.5}
\end{equation}%
is a $(m-1)$-differential form \cite{Go,Ca} on ${\mathbb{R}}^{m}$ with
meanings in $\mathbb{C}.$

Assume now that a pair $(\varphi ,\psi )\in {\mathcal{H}}_{0}^{\ast }\times {%
\mathcal{H}}_{0}\subset {\mathcal{H}}_{-}^{\ast }\times {\mathcal{H}}_{-},$
where, by definition,
\begin{equation}
\begin{array}{c}
{\mathcal{H}}_{0}:=\{\psi (\xi )\in {\mathcal{H}}_{-}:L\psi (\xi )=0,\text{
\ \ \ \ }\psi (\xi )|_{\Gamma }=0,\text{ }\xi \in \Sigma \subset \mathbb{C}%
^{p}\}, \\
{\mathcal{H}}_{0}^{\ast }:=\{\varphi (\eta )\in {\mathcal{H}}_{-}^{\ast
}:L^{\ast }\varphi (\eta )=0,\text{ \ \ }\varphi (\eta )|_{\Gamma }=0,\text{
}\eta \in \Sigma \subset \mathbb{C}^{p}\}%
\end{array}
\label{2.6}
\end{equation}%
with $\Sigma \subset \mathbb{C}^{p}$ being some "spectral" parameter space, $%
\Gamma $ $\subset \mathbb{R}^{m}$ being some $(m-1)$-dimensional
hypersurface piece-wise smoothly imbedded into $\mathbb{R}^{m},$ and ${%
\mathcal{H}}_{-}^{\ast }\supset {\mathcal{H}}^{\ast },$ ${\mathcal{H}}%
_{-}\supset {\mathcal{H}},$ being as before the correspondingly
Hilbert-Schmidt rigged \cite{BS,Be,Na} Hilbert spaces, containing so called
generalized eigenfunctions of the operators $L^{\ast }$ and $L.$ Thereby,
for any pair $(\varphi ,\psi )\in {\mathcal{H}}_{0}^{\ast }\times {\mathcal{H%
}}_{0}$ one gets from (\ref{2.4}) that the differential $m-1$-form $%
Z^{(m-1)}[\varphi ,\psi ]$ is closed in the Grassmann algebra $\Lambda (%
\mathbb{R}^{m};\mathbb{C})$. As a result from the Poincare lemma \cite{Go,Ca}%
, one finds that there exists an ($m-2)$-differential form $\Omega
^{(m-2)}[\varphi ,\psi ]\in \Lambda ^{m-2}(\mathbb{R}^{m};\mathbb{C}),$
semi-linearly depending on ${\mathcal{H}}_{0}^{\ast }\times {\mathcal{H}}%
_{0},$ such that
\begin{equation}
Z^{(m-1)}[\varphi ,\psi ]=d\Omega ^{(m-2)}[\varphi ,\psi ].  \label{2.7}
\end{equation}%
Making use now of the expression (\ref{2.7}), we can get due to the Stokes
theorem \cite{Go,Ca}, that
\begin{equation}
\begin{array}{l}
\int_{{\mathcal{S}}_{+}(\sigma _{x}^{(m-2)},\sigma
_{x_{0}}^{(m-2)})}Z^{(m-1)}[\varphi (\eta ),\psi (\xi )]= \\
\int_{\sigma _{x}^{(m-2)}}\Omega ^{(m-2)}[\varphi (\eta ),\psi (\xi
)]-\int_{\sigma _{x_{0}}^{(m-2)}}\Omega ^{(m-2)}[\varphi (\eta ),\psi (\xi
)]:= \\
\Omega _{x}(\eta ,\xi )-\Omega _{x_{0}}(\eta ,\xi ),%
\end{array}
\label{2.8}
\end{equation}%
for all $(\eta ,\xi )\in \Sigma \times \Sigma ,$ where an $(n-1)$%
-dimensional hypersurface ${\mathcal{S}}_{+}(\sigma _{x}^{(m-2)},\sigma
_{x_{0}}^{(m-2)})\subset {\mathbb{R}}^{m}$ with the boundary $\partial {%
\mathcal{S}}_{+}(\sigma _{x}^{(m-2)},\sigma _{x_{0}}^{(m-2)})=\sigma
_{x}^{(m-2)}-\sigma _{x_{0}}^{(m-2)}$ \ is defined as a film spanned in some
way between two $(m-2)$-dimensional homological nonintersecting each other
cycles $\sigma _{x}^{(m-2)}$ and $\sigma _{x_{0}}^{(m-2)}\subset {\mathbb{R}}%
^{m},$ parametrized, correspondingly, by some arbitrary but fixed points $x$
and $x_{0}\in \mathbb{R}^{m}.$ The quantities $\Omega _{x}(\eta ,\xi )$ and $%
\Omega _{x_{0}}(\eta ,\xi ),$ $(\eta ,\xi )\in \Sigma \times \Sigma ,$
obtained above have to be considered naturally as the corresponding kernels
\cite{BS,Kr,Na} of bounded Hilbert-Schmidt type integral operators $\Omega
_{x},\Omega _{x_{0}}:H\rightarrow H,$ where $H:=L_{2}^{(\rho )}(\Sigma ;%
\mathbb{C})$ is a Hilbert space of \ functions on $\ \Sigma $ measurable
with respect to a finite Borel measure $\rho $ on Borel subsets of $\Sigma ,$
and satisfying the following weak regularity condition
\begin{equation}
lim_{x\rightarrow x_{0}}\Omega (\eta ,\xi )]=\Omega _{x_{0}}(\eta ,\xi )
\label{2.10}
\end{equation}%
for any pair $(\varphi (\eta ),\psi (\xi ))\in {\mathcal{H}}_{0}^{\ast
}\times {\mathcal{H}}_{0},$ $(\eta ,\xi )\in \Sigma \times \Sigma .$

Now we are, similarly to results of \cite{PSS,PSP,GPSP}, in a position to
construct the corresponding pair of spaces $\tilde{\mathcal{H}}_{0}^{\ast }$
and $\tilde{\mathcal{H}}_{0}\subset {\mathcal{H}},$ related with a Delsarte
transformed linear differential operator $\tilde{L}:{\mathcal{H}}%
\longrightarrow {\mathcal{H}}$ and its conjugated expression $\tilde{L}%
^{\ast }:{\mathcal{H}}^{\ast }\longrightarrow {\mathcal{H}}^{\ast },$
\begin{equation}
\tilde{L}(x;\partial ):=\sum_{|\alpha |=0}^{n(\tilde{L})}\tilde{a}_{\alpha
}(x)\frac{\partial ^{|\alpha |}}{\partial x^{\alpha }},  \label{2.11}
\end{equation}%
with coefficients $\tilde{a}_{\alpha }\in {\mathcal{S}}({\mathbb{R}}^{m};End{%
\mathbb{C}}^{N}),$ $\alpha \in \mathbb{Z}_{+}^{m}$ \ is a multi-index, $%
|\alpha |=\overline{0,n(\tilde{L})},$ $x\in {\mathbb{R}}^{m},$ under the
condition $n(\tilde{L})=n(L)\in {\mathbb{Z}}_{+}$ be fixed. Namely, let
closed subspaces $\tilde{\mathcal{H}}_{0}^{\ast }\subset \tilde{\mathcal{H}}%
_{-}^{\ast }$ and $\tilde{\mathcal{H}}_{0}\subset \tilde{\mathcal{H}}_{-}$
be defined as
\begin{equation}
\begin{array}{c}
\tilde{\mathcal{H}}_{0}:=\{\tilde{\psi}(\xi )\in {\mathcal{H}}_{-}:\tilde{%
\psi}(\xi )=\psi (\xi )\cdot \Omega _{x}^{-1}\Omega _{x_{0}}, \\
(\varphi (\eta ),\psi (\xi ))\in {\mathcal{H}}_{0}^{\ast }\times {\mathcal{H}%
}_{0},\text{ }(\eta ,\xi )\in \Sigma \times \Sigma \}, \\
\tilde{\mathcal{H}}_{0}^{\ast }:=\{\tilde{\varphi}(\eta )\in {\mathcal{H}}%
_{-}^{\ast }:\tilde{\varphi}(\eta )=\varphi (\eta )\cdot \Omega
_{x}^{\circledast ,-1}\Omega _{x_{0}}^{\circledast }, \\
(\varphi (\eta ),\psi (\xi ))\in {\mathcal{H}}_{0}^{\ast }\times {\mathcal{H}%
}_{0},\text{ }(\eta ,\xi )\in \Sigma \times \Sigma \}\}.%
\end{array}
\label{2.12}
\end{equation}%
Here, similarly to (\ref{2.8}), we defined the kernels of bounded invertible
integral operators $\Omega _{x}^{\circledast }$ and $\Omega
_{x_{0}}^{\circledast }:H\longrightarrow H$ as follows:
\begin{equation}
\begin{array}{l}
\int_{{\mathcal{S}}_{+}(\sigma _{x}^{(m-2)},\sigma _{x_{0}}^{(m-2)})}\bar{Z}%
^{(m-1),\intercal }[\varphi (\eta ),\psi (\xi )] \\
=\int_{\sigma _{x}^{(m-2)}}\bar{\Omega}^{(m-2),\intercal }[\varphi (\eta
),\psi (\xi )]-\int_{\sigma _{x_{0}}^{(m-2)}}\bar{\Omega}^{(m-2),\intercal
}[\varphi (\eta ),\psi (\xi )] \\
:=\Omega _{x}^{\circledast }(\eta ,\xi )-\Omega _{x_{0}}^{\circledast }(\eta
,\xi )%
\end{array}
\notag
\end{equation}%
for all $(\eta ,\xi )\in \Sigma \times \Sigma ,$ where homological $(m-2)$%
-cycles $\sigma _{x}^{(m-2)}$ and $\sigma _{x_{0}}^{(m-2)}\subset {\mathbb{R}%
}^{m}$ are the same as taken above. Thereby, making use of the classical
method of variation of constants as in \cite{PSP,SP,PSS}, one gets easily
from (2.11) that for any $(\varphi (\eta ),\psi (\xi ))\in {\mathcal{H}}%
_{0}^{\ast }\times {\mathcal{H}}_{0},$ $(\eta ,\xi )\in \Sigma \times \Sigma
,$
\begin{equation}
\tilde{\psi}(\xi )=\mathbf{\Omega }_{+}\psi (\xi ),\qquad \tilde{\varphi}%
(\eta )=\mathbf{\Omega }_{+}^{\circledast }\varphi (\eta ),  \label{2.14}
\end{equation}%
where integral expressions
\begin{equation}
\begin{array}{l}
\mathbf{\Omega }_{+}:=\mathbf{1}-\int_{\Sigma }d\rho (\xi )\int_{\Sigma
}d\rho (\eta )\tilde{\psi}(\xi )\Omega _{x_{0}}^{-1}(\xi ,\eta )\int_{{%
\mathcal{S}}_{+}(\sigma _{x}^{(m-2)},\sigma
_{x_{0}}^{(m-2)})}Z^{(m-1)}[\varphi (\eta ),\cdot ], \\
\mathbf{\Omega }_{+}^{\circledast }:=\mathbf{1}-\int_{\Sigma }d\rho (\xi
)\int_{\Sigma }d\rho (\eta )\tilde{\varphi}(\eta )\Omega
_{x_{0}}^{\circledast ,-1}(\xi ,\eta )\int_{{\mathcal{S}}_{+}(\sigma
_{x}^{(m-2)},\sigma _{x_{0}}^{(m-2)})}\bar{Z}^{(m-1),\intercal }[\cdot ,\psi
(\xi )]%
\end{array}
\label{2.15}
\end{equation}%
are bounded Delsarte transmutation operators of Volterra type defined,
correspondingly, on the whole spaces ${\mathcal{H}}$ and ${\mathcal{H}}%
^{\ast }.$

Now, based on operator expressions (\ref{2.15}) and the definition (\ref{1.2}%
), one gets easily the expressions for Delsarte transformed operators $%
\tilde{L}$ and $\tilde{L}^{\ast }:$
\begin{equation}
\begin{array}{l}
\tilde{L}=\mathbf{\Omega }_{+}L\mathbf{\Omega }_{+}^{-1}=L+[\mathbf{\Omega }%
_{+},L]\mathbf{\Omega }_{+}^{-1}, \\
\tilde{L}^{\ast }=\mathbf{\Omega }_{+}^{\circledast }L\mathbf{\Omega }%
_{+}^{\circledast ,-1}=L^{\ast }+[\mathbf{\Omega }_{+}^{\circledast
},L^{\ast }]\mathbf{\Omega }_{+}^{\circledast ,-1}.%
\end{array}
\label{2.16}
\end{equation}%
Note also here that the transformations like (\ref{2.14}) were for
one-dimensional case in detail studied in \cite{Ma,Na,LS}. They satisfy
evidently the following easily found conditions:
\begin{equation}
\tilde{L}\tilde{\psi}=0,\qquad \tilde{L}^{\ast }\tilde{\varphi}=0
\label{2.17}
\end{equation}%
for any pair $(\tilde{\varphi},\tilde{\psi})\in \tilde{\mathcal{H}}%
_{0}^{\ast }\times \tilde{\mathcal{H}}_{0},$ which can be specified by
constraints
\begin{equation}
\tilde{\psi}|_{\tilde{\Gamma}}=0,\qquad \tilde{\varphi}|_{\tilde{\Gamma}%
^{\ast }}=0  \label{2.18}
\end{equation}%
for some hypersurface $\tilde{\Gamma}$ $\subset {\mathbb{R}}^{m},$ related
with the previously chosen hypersurface ${\Gamma }$ $\subset {\mathbb{R}}%
^{m} $ and the homological pair of $(m-2)$-dimensional cycles $\sigma
_{x}^{(m-2)} $ and $\sigma _{x_{0}}^{(m-2)}$ $\subset {\mathbb{R}}^{m}.$
Thereby, the closed subspaces $\tilde{\mathcal{H}}_{0}$ and $\tilde{\mathcal{%
H}}_{0}^{\ast }$ can be re-defined similarly to (\ref{2.6}) :
\begin{equation}
\begin{array}{c}
{\mathcal{\tilde{H}}}_{0}:=\{\tilde{\psi}(\xi )\in {\mathcal{H}}_{-}:\tilde{L%
}\psi (\xi )=0,\text{ \ \ \ \ }\tilde{\psi}(\xi )|_{\tilde{\Gamma}}=0,\text{
}\xi \in \Sigma \subset \mathbb{C}^{p}\}, \\
{\mathcal{\tilde{H}}}_{0}^{\ast }:=\{\tilde{\varphi}(\eta )\in {\mathcal{H}}%
_{-}^{\ast }:\tilde{L}^{\ast }\tilde{\varphi}(\eta )=0,\text{ \ \ }\tilde{%
\varphi}(\eta )|_{\tilde{\Gamma}}=0,\text{ }\eta \in \Sigma \subset \mathbb{C%
}^{p}\}%
\end{array}
\label{2.19}
\end{equation}%
Moreover, the following lemma, based on a pseudo-differential operators
technique from \cite{BS,Na,SP} , holds.

\begin{lemma}
The Delsarte transformed operators (\ref{2.15}) by means of transmutation
operators (\ref{2.14}) are differential too if the starting operator $L:{%
\mathcal{H}}\longrightarrow {\mathcal{H}}$ \ was taken differential.
\end{lemma}

As a simple consequence of the structure of the constructed above Delsarte
transformed operators (\ref{2.15}) one states that for any pair $(\tilde{%
\varphi},\tilde{\psi})\in \tilde{\mathcal{H}}_{0}^{\ast }\times \tilde{%
\mathcal{H}}_{0}.$ The following differential forms equality holds:
\begin{equation}
\tilde{Z}^{(m-1)}[\tilde{\varphi},\tilde{\psi}]=d\tilde{\Omega}^{(m-2)}[%
\tilde{\varphi},\tilde{\psi}],  \label{2.20}
\end{equation}%
where, by definition, a pair $(\tilde{\varphi},\tilde{\psi})\in \tilde{%
\mathcal{H}}_{0}^{\ast }\times \tilde{\mathcal{H}}_{0}$ is fixed and the
equality
\begin{equation}
\left( <\tilde{L}^{\ast }\tilde{\varphi},\tilde{\psi}>-<\tilde{\varphi},%
\tilde{L}\tilde{\psi}>\right) dx=d\tilde{Z}^{(m-1)}[\tilde{\varphi},\tilde{%
\psi}]  \label{2.21}
\end{equation}%
holds. The equality (\ref{2.19}) makes it possible to construct the
corresponding kernels
\begin{equation}
\begin{array}{c}
\tilde{\Omega}_{x}(\eta ,\xi ):=\int_{\sigma _{x}^{(m-2)}}\tilde{\Omega}%
^{(m-2)}[\tilde{\varphi}(\eta ),\tilde{\psi}(\xi )], \\
\tilde{\Omega}_{x_{0}}(\eta ,\xi )]:=\int_{\sigma _{x_{0}}^{(m-2)}}\tilde{%
\Omega}^{(m-2)}[\tilde{\varphi}(\eta ),\tilde{\psi}(\xi )]%
\end{array}
\label{2.22}
\end{equation}%
of bounded integral invertible Hilbert-Schmidt operators $\tilde{\Omega}_{x},%
\tilde{\Omega}_{x_{0}}:H\rightarrow H,$ and corresponding kernels
\begin{equation}
\begin{array}{c}
\tilde{\Omega}_{x}^{\circledast }(\eta ,\xi ):=\int_{\sigma _{x}^{(m-2)}}%
\bar{\tilde{\Omega}}^{(m-2),\intercal }[\tilde{\varphi}(\eta ),\tilde{\psi}%
(\xi )], \\
\tilde{\Omega}_{x_{0}}^{\circledast }(\eta ,\xi ):=\int_{\sigma
_{x_{0}}^{(m-2)}}\bar{\tilde{\Omega}}^{(m-2,\intercal )}[\tilde{\varphi}%
(\eta ),\tilde{\psi}(\xi )]%
\end{array}
\label{2.23}
\end{equation}%
of bounded integral invertible Hilbert-Schmidt operators $\tilde{\Omega}_{x},%
\tilde{\Omega}_{x_{0}}:H^{\ast }\rightarrow H^{\ast }.$ Then the following
equalities hold for all mutually related pairs $(\varphi ,\psi )\in {%
\mathcal{H}}_{0}^{\ast }\times {\mathcal{H}}_{0}$ and $(\tilde{\varphi},%
\tilde{\psi})\in \tilde{\mathcal{H}}_{0}^{\ast }\times \tilde{\mathcal{H}}%
_{0}:$
\begin{equation}
\begin{array}{c}
\psi (\xi )=\tilde{\psi}(\xi )\cdot \tilde{\Omega}_{x}^{-1}\tilde{\Omega}%
_{x_{0}}, \\
\varphi (\eta )=\tilde{\varphi}(\eta )\cdot \tilde{\Omega}_{x}^{\circledast
,-1}\tilde{\Omega}_{x_{0}}^{\circledast }%
\end{array}
\label{2.24}
\end{equation}%
where $(\eta ,\xi )\in \Sigma \times \Sigma .$ Thus, based on the symmetry
property between relations (\ref{2.11}) and (\ref{2.23}), one easily finds
from expression (\ref{2.24}), that expressions
\begin{equation}
\begin{array}{l}
\mathbf{\Omega }_{+}^{-1}:=\mathbf{1}-\int_{\Sigma }d\rho (\xi )\int_{\Sigma
}d\rho (\eta )\psi (\xi )\tilde{\Omega}_{x_{0}}^{-1}(\xi ,\eta )\int_{{%
\mathcal{S}}_{+}(\sigma _{x}^{(m-2)},\sigma _{x_{0}}^{(m-2)})}\tilde{Z}%
^{(m-1)}[\tilde{\varphi}(\eta ),\cdot ], \\
\mathbf{\Omega }_{+}^{\circledast ,-1}:=\mathbf{1}-\int_{\Sigma }d\rho (\xi
)\int_{\Sigma }d\rho (\eta )\varphi (\eta )\tilde{\Omega}_{x_{0}}^{%
\circledast ,-1}(\xi ,\eta )\int_{{\mathcal{S}}_{+}(\sigma
_{x}^{(m-2)},\sigma _{x_{0}}^{(m-2)})}\overset{\_}{\tilde{Z}}%
^{(m-1),\intercal }[\cdot ,\tilde{\psi}(\xi )]%
\end{array}
\label{2.25}
\end{equation}%
for some homological $(m-2)$-dimensional cycles $\tilde{\sigma}_{x}^{(m-2)},%
\tilde{\sigma}_{x_{0}}^{(m-2)}\subset \mathbb{R}^{m}$ are inverse to (2.14)
Delsarte transmutation integral operators of\ \ Volterra type, satisfying
the following relationships:
\begin{equation}
\psi (\xi )=\mathbf{\Omega }_{+}^{-1}\cdot \tilde{\psi}(\xi ),\qquad \varphi
(\eta )=\mathbf{\Omega }_{+}^{\ast ,-1}\cdot \tilde{\varphi}(\eta )
\label{2.26}
\end{equation}%
for all arbitrary but fixed pairs of functions $(\varphi (\eta ),\psi (\xi
))\in {\mathcal{H}}_{0}^{\ast }\times {\mathcal{H}}_{0}$ and $(\tilde{\varphi%
}(\eta ),\tilde{\psi}(\xi ))\in \tilde{\mathcal{H}}_{0}^{\ast }\times \tilde{%
\mathcal{H}}_{0},$ $(\eta ,\xi )\in \Sigma \times \Sigma .$ Thus, one can
formulate the following characterizing the constructed Delsarte
transmutation operators theorem.

\begin{theorem}
Let a matrix multi-dimensional differential operator (\ref{2.1}) acting in a
Hilbert space ${\mathcal{H}}=L_{2}(\mathbb{R}^{m};\mathbb{C}^{N})$ and its
formally adjoint operator (\ref{2.2}) acting in a Hilbert space ${\mathcal{H}%
}^{\ast }=L_{2}^{\ast }(\mathbb{R}^{m};\mathbb{C}^{N}),$ possess,
correspondingly, a pair of closed spaces ${\mathcal{H}}_{0}$ and ${\mathcal{H%
}}_{0}^{\ast }$ (\ref{2.6}) of their generalized kernel eigenfunctions
parametrized by some set $\Sigma \subset \mathbf{C}^{p}.$ Then there exist
bounded invertible Delsarte transmutation integral operators $\mathbf{\Omega
}_{+}:{\mathcal{H}}\longrightarrow {\mathcal{H}}$ \ and $\Omega
_{+}^{\circledast }:{\mathcal{H}}^{\ast }\longrightarrow {\mathcal{H}}^{\ast
},$ such that for this pair $({\mathcal{H}}_{0},{\mathcal{H}}_{0}^{\ast })$ (%
\ref{2.6}) of closed subspaces (2.6) and their dual ones (\ref{2.19}) the
corresponding bounded invertible mappings (\ref{2.15}) \ $\mathbf{\Omega }%
_{+}:{\mathcal{H}}_{0}\rightarrow \tilde{\mathcal{H}}_{0}$ and $\mathbf{%
\Omega }_{+}^{\circledast }:{\mathcal{H}}_{0}^{\ast }\rightarrow \tilde{%
\mathcal{H}}_{0}^{\ast }$ are compatibly defined. Moreover, the operator
expressions (\ref{2.16}) are also differential, acting in the corresponding
spaces ${\mathcal{H}}$ \ and ${\mathcal{H}}^{\ast }.$
\end{theorem}

The revealed above structure of the Delsarte transmutation operators (\ref%
{2.15}) makes it possible to understand more deeply their properties by
means of deriving new integral equations being multi-dimensional analogs of
the well known Gelfand-Levitan-Marchenko equations \cite{Ma, LS,Na,Ni}, that
will be a topic of the next chapter.

\section{Multi-dimensional Gelfand-Levitan-Marchenko type integral equations}

Investigating the inverse scattering problem for a three-dimensional
perturbed Laplace operator
\begin{equation}
L(x;\partial )=-\sum_{j=1}^{3}\frac{\partial ^{2}}{\partial x_{j}^{2}}+q(x),
\label{3.1}
\end{equation}%
with $q\in W_{2}^{2}(\mathbb{R}^{3}),$ $x\in \mathbb{R}^{3},$ in the Hilbert
space ${\mathcal{H}}=L_{2}(\mathbb{R}^{3};\mathbb{C}),$ L.D. Faddeev in \cite%
{Fa} has suggested some approach to studying the structure of the
corresponding Delsarte transmutation operators $\mathbf{\Omega }_{\gamma }:$
${\mathcal{H\rightarrow H}}$ of Volterra type, based on a priori chosen
hypersurfaces $S_{\pm \gamma }^{(x)}=\{y\in \mathbb{R}^{3}:<y-x,\pm \gamma >$
$>0\},$ parametrized by unity vectors $\gamma \in {\mathbb{S}}^{2},$ where ${%
\mathbb{S}}^{2}\subset \mathbb{R}^{3}$ is the standard two-dimensional
sphere imbedded into $\mathbb{R}^{3}.$ Making use of these Delsarte
transmutation operators of Volterra type, in \cite{Fa} \ there was derived
some three-dimensional analog of the integral GLM-equation, whose solution
gives rise to the kernel of the corresponding Delsarte transmutation
operator for (\ref{3.1}) . But the important two problems related with this
approach were not discussed in detail: the first one concerns the question
whether the Delsarte transformed operator $\tilde{L}=\mathbf{\Omega }%
_{\gamma }L\mathbf{\hat{\Omega}}_{\gamma }^{-1}$ is also a differential
operator of Laplace type, and the second one concerns the question of
existing Delsarte transmutation operators in the Faddeev form.

Below we will study our multi-dimensional Delsarte transmutation operators
(2.14), parametrized by a hypersurface $S_{+}(\sigma _{x}^{(m-2)},\sigma
_{x_{0}}^{(m-2)})$ piecewise smoothly imbedded into ${\mathbb{R}}^{m}.$

Consider now some $(m-2)$-dimensional homological cycles $\sigma
_{x}^{(m-2)} $ and $\sigma _{x_{0}}^{(m-2)}\subset {\mathbb{R}}^{m}$ and two
$(m-1)$-dimensional smooth hypersurfaces
\begin{equation*}
S_{+}(\sigma _{x}^{(m-2)},\sigma _{x_{0}}^{(m-2)}),\text{ \ \ }S_{-}(\sigma
_{x}^{(m-2)},\sigma _{x_{0}}^{(m-2)})
\end{equation*}%
spanned between them in such a way that the whole hypersurface ${\mathcal{S}}%
_{+}(\sigma _{x}^{(m-2)},\sigma _{x_{0}}^{(m-2)})$ $\cup {\mathcal{S}}%
_{-}(\sigma _{x}^{(m-2)},\sigma _{x_{0}}^{(m-2)})$ is closed. Then similarly
to the construction of Chapter 3 one can naturally define two pairs of
Delsarte transmutation operators for a given pair of multi-dimensional
differential operators (\ref{2.1}) and (\ref{2.10}), namely operators $%
\mathbf{\Omega }_{+}:{\mathcal{H}}\rightleftharpoons {\mathcal{H}},$ $%
\mathbf{\Omega }_{+}^{\circledast }:{\mathcal{H}}^{\ast }\rightleftharpoons {%
\mathcal{H}}^{\ast },$ defined by (\ref{2.15}), and operators $\mathbf{%
\Omega }_{-}^{\circledast }:{\mathcal{H}}^{\ast }\rightleftharpoons {%
\mathcal{H}}^{\ast },$ $\mathbf{\Omega }_{-}^{\circledast }:{\mathcal{H}}%
^{\ast }\rightleftharpoons {\mathcal{H}}^{\ast },$ where, by definition,%
\begin{equation}
\begin{array}{l}
\mathbf{\Omega }_{-}:=\mathbf{1}-\int_{\Sigma }d\rho (\xi )\int_{\Sigma
}d\rho (\eta )\tilde{\psi}(\xi )\Omega _{x_{0}}^{-1}(\xi ,\eta )\int_{{%
\mathcal{S}}_{-}(\sigma _{x}^{(m-2)},\sigma
_{x_{0}}^{(m-2)})}Z^{(m-1)}[\varphi (\eta ),\cdot ], \\
\mathbf{\Omega }_{-}^{\circledast }:=\mathbf{1}-\int_{\Sigma }d\rho (\xi
)\int_{\Sigma }d\rho (\eta )\tilde{\varphi}(\eta )\Omega
_{x_{0}}^{\circledast ,-1}(\xi ,\eta )\int_{{\mathcal{S}}_{-}(\sigma
_{x}^{(m-2)},\sigma _{x_{0}}^{(m-2)})}\bar{Z}^{(m-1),\intercal }[\cdot ,\psi
(\xi )]%
\end{array}
\label{3.2}
\end{equation}%
\ Subject to the Delsarte transmutation operators (\ref{2.14}) ) and (\ref%
{3.2}) the following operator relationships

\begin{equation*}
\tilde{L}=\mathbf{\Omega }_{\pm }L\mathbf{\Omega }_{\pm }^{-1},\text{ \ \ \ }%
\mathbf{\Omega }_{\pm }^{\circledast }L_{\pm }^{\circledast }\mathbf{\Omega }%
_{\pm }^{\circledast ,-1}=\tilde{L}^{\ast }
\end{equation*}%
hold. As in theory of classical GLM-equations \cite{Ma,LS,Na}, we can now
construct linear integral operators of Fredholm type $\mathbf{\Phi }:{%
\mathcal{H}}\rightarrow {\mathcal{H}},$ $\mathbf{\Phi }^{\circledast }:{%
\mathcal{H}}^{\ast }\rightarrow {\mathcal{H}}^{\ast }$ of \ Fredholm type,
such that
\begin{equation}
\mathbf{1}+\mathbf{\Phi }:=\mathbf{\Omega }_{+}^{-1}\cdot \mathbf{\Omega }%
_{-},\qquad \mathbf{1}+\mathbf{\Phi }^{\circledast }:=\mathbf{\Omega }%
_{+}^{\circledast ,-1}\cdot \mathbf{\Omega }_{-}^{\circledast }.  \label{3.3}
\end{equation}%
Making use of the expressions (\ref{3.3}), one easily gets a pair of linear
integral GLM-equations:
\begin{equation}
\mathbf{\Omega }_{+}\cdot (\mathbf{1}+\mathbf{\Phi })=\mathbf{\Omega }%
_{-},\qquad \mathbf{\Omega }_{+}^{\circledast }\cdot (\mathbf{1}+\mathbf{%
\Phi }^{\circledast })=\mathbf{\Omega }_{-}^{\circledast },  \label{3.4}
\end{equation}%
whose solution is a pair of the corresponding Volterra type kernels for the
Delsarte transmutation operators $\mathbf{\Omega }_{+}$ and $\mathbf{\Omega }%
_{+}^{\circledast }.$ Thus, the problem of constructing Delsarte
transmutation operators for a given pair of differential operators (\ref{2.1}%
) and (\ref{2.10}) is reduced to that of describing a suitable class of
linear Fredholm type operators (\ref{3.3}) in the Hilbert space $\mathcal{H}%
, $ satisfying the following natural conditions: operators $(\mathbf{1}+%
\mathbf{\Phi }):{\mathcal{H}}\longrightarrow {\mathcal{H}}$ \ and $(\mathbf{1%
}+\mathbf{\Phi }^{\circledast }):{\mathcal{H}}^{\ast }\longrightarrow {%
\mathcal{H}}^{\ast }$ are onto, bounded and invertible and, moreover,
\begin{equation}
(\mathbf{1}+\mathbf{\Phi })L=L(\mathbf{1}+\mathbf{\Phi }),\text{ \ }(\mathbf{%
1}+\mathbf{\Phi }^{\circledast })L^{\ast }=L^{\ast }(\mathbf{1}+\mathbf{\Phi
}^{\circledast })  \label{3.5}
\end{equation}%
due to (\ref{3.4}) and (\ref{2.15}) . This problem is very important for the
theory devised here could be effectively applied to studying diverse
spectral properties of a given pair of Delsarte transformed differential
operators (\ref{2.1}) and (\ref{2.10}) and is planned to be studied in
detail in another place.

\section{The structure of Delsarte transmutation operators for affine
pencils of multidimensional differential expressions}

\setcounter{equation}{0}Consider in the Hilbert space ${\mathcal{H}}=L_{2}(%
\mathbb{R}^{m};\mathbb{C}^{N})$ an affine polynomial in $\lambda \in {%
\mathbb{C}}$ pencil of multi-dimensional differential operators
\begin{equation}
L(x;\partial |\lambda ):=\sum_{i=0}^{r(L)}\lambda ^{i}L_{i}(x;\partial ),
\label{4.1}
\end{equation}%
where $x\in \mathbb{R}^{m},$ $ordL_{i}(x;\partial )=n_{i}\in {\mathbb{Z}}%
_{+},$ $i=\overline{1,r(L)},$ the order $r(L)\in {\mathbb{Z}}_{+}$ is fixed
and
\begin{equation}
L_{i}(x;\partial ):=\sum_{|\alpha _{i}|=0}^{n_{i}}a_{i,\alpha _{i}}(x)\frac{%
\partial ^{|\alpha _{i}|}}{\partial x^{\alpha _{i}}}  \label{4.2}
\end{equation}%
are differential expressions with smooth coefficients $a_{i,\alpha _{i}}\in {%
\mathcal{S}}(\mathbb{R}^{m};End\mathbb{C}^{N}),$ $i=\overline{1,r(L)}.$ The
pencil (\ref{4.1}) can be, in particular, characterized by its spectrum
\begin{equation}
\sigma (L)=\{\lambda \in \mathbb{C}:\exists \text{ }\psi (x;\lambda )\in {%
\mathcal{H}}_{-},\text{ \ }L(x;\partial |\lambda )\psi (x;\lambda )=0\}.
\label{4.3}
\end{equation}%
As was demonstrated in \cite{DS}. the transformations of pencil (\ref{4.1})
which preserve a part of the spectrum $\sigma (L)$ and simultaneously change
in a prescribed way the rest of the spectrum (so called an assignment
spectrum problem \cite{DS} ) are of very importance for feedback control
theory and its applications in different fields of mechanics.

We will try here to interpret these "spectrum assignment" transformations as
ones of Delsarte transmutation type, satisfying some additional special
conditions. Thus, we look for such a transformation $\mathbf{\Omega }:{%
\mathcal{H}}\longrightarrow {\mathcal{H}}$ of the pencil (4.1) into a
similar pencil
\begin{equation}
\tilde{L}(x;\partial |\lambda )=\sum_{i=1}^{r(L)}\lambda ^{i}\tilde{L}%
_{i}(x;\partial ),\qquad \tilde{L}_{i}(x;\partial ):=\sum_{|\alpha
_{i}|=0}^{n_{i}}\tilde{a}_{i,\alpha _{i}}(x)\frac{\partial ^{|\alpha _{i}|}}{%
\partial x^{\alpha _{i}}}  \label{4.4}
\end{equation}%
with $\tilde{a}_{i,\alpha _{i}}\in S(\mathbb{R}^{m};End\mathbb{C}^{N}),$ $i=%
\overline{1,r(L)},$ $\lambda \in {\mathbb{C}},$ of the same polynomial and
differential orders, that
\begin{equation}
\tilde{L}=L+[\mathbf{\Omega },L]\mathbf{\Omega }^{-1}=\mathbf{\Omega }L%
\mathbf{\Omega }^{-1}.  \label{4.5}
\end{equation}

For such an operator $\mathbf{\Omega }:{\mathcal{H}}\rightarrow {\mathcal{H}}
$ to be constructed, we suggest to extend the polynomial pencil of
differential operators (\ref{4.1}) to a pure differential operator $L_{\tau
}:=L(x;\partial |\partial /\partial \tau ),$ $x\in {\mathbb{R}}^{m},$ $\tau
\in {\mathbb{R}},$ with ${\mathbb{R}}\ni \tau $-independent coefficients and
acting suitably in the parametric functional space ${\mathcal{H}}_{(\tau
)}:=L_{1}({\mathbb{R}}_{\tau };{\mathcal{H}}).$ Thereby we have come at the
same situation, which was studied before in \cite{GPSP}. For completeness,
we shall deliver here a short derivation of the corresponding affine
expression for the Delsarte transmutation operator $\mathbf{\Omega }:{%
\mathcal{H}}\longrightarrow {\mathcal{H}}.$

Let a pair of functions $(\varphi _{(\tau )},\psi _{(\tau )})\in {\mathcal{H}%
}_{(\tau )}^{\ast }\times {\mathcal{H}}_{(\tau )}$ be arbitrary and consider
the following semi-linear scalar form on ${\mathcal{H}}_{(\tau )}^{\ast
}\times {\mathcal{H}}_{(\tau )}$ :
\begin{equation}
(\varphi _{(\tau )},\psi _{(\tau )}):=\int_{{\mathbb{R}}_{\tau }}d\tau \int_{%
\mathbb{R}^{m}}dx{\bar{\varphi}_{(\tau )}}^{\intercal }(x)\psi _{(\tau )}(x).
\label{4.6}
\end{equation}%
Then subject to the internal semi-linear form $<\cdot ,\cdot >$ on $\mathbb{C%
}^{N}\mathbb{\times C}^{N}$ one can write down for the operator $L_{(\tau )}:%
{\mathcal{H}}_{(\tau )}\longrightarrow {\mathcal{H}}_{(\tau )}$ and any pair
$(\varphi _{(\tau )},\psi _{(\tau )})\in {\mathcal{H}}_{(\tau )}^{\ast
}\times {\mathcal{H}}_{(\tau )}$ the following Lagrangian identity:
\begin{equation}
\left[ <L_{(\tau )}^{\ast }\varphi _{(\tau )},\psi _{(\tau )}>-<\varphi
_{(\tau )},L_{(\tau )}\psi _{(\tau )}>\right] d\tau \wedge dx=dZ_{(\tau
)}^{(m)}[\varphi ,\psi ],  \label{4.7}
\end{equation}%
where $Z_{(\tau )}^{(m)}[\varphi ,\psi ]\in \Lambda ^{m}({\mathbb{R}}_{\tau
}\times {\mathbb{R}}^{m};\mathbb{C})$ is the corresponding differential $m$%
-form with values in $\mathbb{C},$ parametrically depending on $\tau \in {%
\mathbb{R}}.$ Thus, for defining the closed subspaces ${\mathcal{H}}_{(\tau
),0}^{\ast }\subset {\mathcal{H}}_{(\tau ),-}^{\ast }$ one can write down,
correspondingly, the following expressions:
\begin{equation}
\begin{array}{c}
{\mathcal{H}}_{(\tau ),0}:=\{\psi _{(\tau )}(\xi )\in {\mathcal{H}}_{(\tau
),-}:L_{(\tau )}\psi _{(\tau )}(\xi )=0, \\
\tau \in \mathbb{R},\text{ }\psi _{(\tau )}(\xi )|_{\Gamma }=0,\text{ }\xi
\in \Sigma \subset \mathbb{C}^{p}\}, \\
{\mathcal{H}}_{(\tau ),0}^{\ast }:=\{\varphi _{(\tau )}(\eta )(\in {\mathcal{%
H}}_{(\tau ),-}^{\ast }:L_{(\tau )}^{\ast }\varphi _{(\tau )}(\eta )=0, \\
\tau \in {\mathbb{R}},\text{ \ }\varphi _{(\tau )}(\eta )|_{\Gamma }=0,\text{
}\eta \in \Sigma \subset \mathbb{C}^{p}\},%
\end{array}
\label{4.8}
\end{equation}%
where ${\Gamma }\subset ${$\mathbb{R}$}$^{m}$ is some piecewise smooth
boundary hypersurfaces in {$\mathbb{R}$}$^{m}.$ Similar expressions one can
also write down for the Delsarte transformed operator expression $\tilde{L}:{%
\mathcal{H}}_{(\tau )}\longrightarrow {\mathcal{H}}_{(\tau )}:$
\begin{equation}
\begin{array}{c}
\tilde{\mathcal{H}}_{(\tau ),0}:=\{\tilde{\psi}_{(\tau )}(\xi )\in {\mathcal{%
H}}_{(\tau ),-}:\tilde{L}_{(\tau )}\tilde{\psi}_{(\tau )}(\xi )=0, \\
\tau \in {\mathbb{R}},\text{ }\tilde{\psi}_{(\tau )}(\xi )|_{\tilde{\Gamma}%
}=0,\text{ }\xi \in \Sigma \subset \mathbb{C}^{p}\}, \\
\tilde{\mathcal{H}}_{(\tau ),0}^{\ast }:=\{\tilde{\varphi}_{(\tau )}(\eta
)\in {\mathcal{H}}_{(\tau ),-}^{\ast }:\tilde{L}_{(\tau )}^{\ast }\tilde{%
\varphi}_{(\tau )}(\eta )=0, \\
\tau \in {\mathbb{R}},\text{ \ }\tilde{\varphi}_{(\tau )}(\eta )|_{\tilde{%
\Gamma}}=0,\text{ }\eta \in \Sigma \subset \mathbb{C}^{p}\},%
\end{array}
\label{4.9}
\end{equation}%
where $\tilde{\Gamma}\subset ${$\mathbb{R}$}$^{3}$ is some piecewise smooth
boundary hypersurface in {$\mathbb{R}$}$^{m}.$

Making use of the expressions (\ref{4.6}) and (\ref{4.7}), we easily find
that the differential $m$-form $Z_{(\tau )}^{(m)}[\varphi ,\psi ]\in \Lambda
^{m}({\mathbb{R}}_{\tau }\times {\mathbb{R}}^{m};\mathbb{C})$ is exact for
any pair $(\varphi ,\psi )\in {\mathcal{H}}_{(\tau ),0}^{\ast }\times {%
\mathcal{H}}_{(\tau ),0}.$ This means due to the Poincare lemma \cite{Go,Ca}%
, that there exists a differential $(m-1)$-form $\Omega _{\tau
}^{(m-1)}[\varphi ,\psi ]\in \Lambda ^{m-1}({\mathbb{R}}_{\tau }\times {%
\mathbb{R}}^{m};\mathbb{C}),$ such that
\begin{equation}
Z_{(\tau )}^{(m)}[\varphi ,\psi ]=d\Omega _{\tau }^{(m-1)}[\varphi ,\psi ]
\label{4.10}
\end{equation}%
for all pairs $(\varphi _{(\tau )},\psi _{(\tau )})\in {\mathcal{H}}_{(\tau
),0}^{\ast }\times {\mathcal{H}}_{(\tau ),0}.$ Now we are in a starting
position for defining the corresponding Delsarte transmutation operators $%
\mathbf{\Omega }_{(\tau )}:{\mathcal{H}}_{(\tau ),0}\rightarrow \tilde{%
\mathcal{H}}_{(\tau ),0}$ and $\mathbf{\Omega }_{(\tau )}^{\circledast }:{%
\mathcal{H}}_{(\tau ),0}^{\ast }\rightarrow \tilde{\mathcal{H}}_{(\tau
),0}^{\ast }:$
\begin{equation}
\begin{array}{l}
\tilde{\psi}_{(\tau )}(\xi )=\mathbf{\Omega }_{(\tau )}\cdot \psi _{(\tau
)}(\xi ):=\psi _{(\tau )}(\xi )\cdot \Omega _{(x,\tau )}^{-1}\Omega
_{(x_{0},\tau )}= \\
(\mathbf{1}-\tilde{\psi}_{(\tau )}\Omega _{(x_{0},\tau )}^{-1}\int_{{%
\mathcal{S}}(\sigma _{x}^{(m-1)},\sigma _{x_{0}}^{(m-1)})}Z_{(\tau
)}^{(m)}[\varphi _{(\tau )},\cdot ])\psi _{(\tau )}(\xi ), \\
\tilde{\varphi}_{(\tau )}(\eta )=\mathbf{\Omega }_{(\tau )}^{\circledast
}\cdot \varphi _{(\tau )}(\eta ):=\varphi _{(\tau )}(\eta )\cdot \Omega
_{(x,\tau )}^{\circledast ,-1}\Omega _{(x_{0},\tau )}^{\circledast }= \\
(\mathbf{1}-\tilde{\varphi}_{(\tau )}\Omega _{(x_{0},\tau )}^{\circledast
,-1}\int_{{\mathcal{S}}(\sigma ,_{x}^{(m-1)},\sigma _{x_{0}}^{(m-1)})}\bar{Z}%
_{(\tau )}^{(m),\intercal }[\cdot ,\psi _{(\tau )}])\varphi _{(\tau )}(\eta
),%
\end{array}
\label{4.11}
\end{equation}%
Here $(\varphi _{(\tau )}(\eta ),\psi _{(\tau )}(\xi ))\in {\mathcal{H}}%
_{(\tau ),0}^{\ast }\times {\mathcal{H}}_{(\tau ),0},$ $\xi ,\eta \in \Sigma
,$ and, due to (\ref{4.9}), for kernels $\Omega _{(x,\tau )}(\eta ,\xi ),$ $%
\Omega _{(x_{0},\tau )}(\eta ,\xi )$ $\in H\otimes H$ $\ $and $\ \Omega
_{(x,\tau )}^{\circledast }(\eta ,\xi ),$ $\Omega _{(x_{0},\tau
)}^{\circledast }(\eta ,\xi )\in H^{\ast }\otimes H^{\ast }$ of the
corresponding integral operators $\Omega _{(x,\tau )},$ $\Omega
_{(x_{0},\tau )}:H\rightarrow H$ and $\ \Omega _{(x,\tau )}^{\circledast },$
$\Omega _{(x_{0},\tau )}^{\circledast }:H^{\ast }\rightarrow H^{\ast }$ one
has
\begin{equation}
\begin{array}{l}
\int_{{\mathcal{S}}(\sigma _{x}^{(m-1)},\sigma _{x_{0}}^{(m-1)})}Z_{(\tau
)}^{(m)}[\varphi _{(\tau )}(\eta ),\psi _{(\tau )}(\xi )]=\int_{\sigma
_{x}^{(m-1)}}\Omega _{(\tau )}^{(m-1)}[\varphi _{(\tau )}(\eta ),\psi
_{(\tau )}(\xi )] \\
-\int_{\sigma _{x_{0}}^{(m-1)}}\Omega _{(\tau )}^{(m-1)}[\varphi _{(\tau
)}(\eta ),\psi _{(\tau )}(\xi )]:=\Omega _{(x,\tau )}(\eta ,\xi )-\Omega
_{(x_{0},\tau )}(\eta ,\xi ),%
\end{array}
\label{4.12}
\end{equation}%
\begin{equation*}
\begin{array}{l}
\int_{{\mathcal{S}}(\sigma ,_{x}^{(m-1)},\sigma _{x_{0}}^{(m-1)})}\bar{Z}%
_{(\tau )}^{(m),\intercal }[\varphi _{(\tau )}(\eta ),\psi _{(\tau )}(\xi
)]=\int_{\sigma _{x}^{(m-1)}}\bar{\Omega}_{(\tau )}^{(m-1),\intercal
}[\varphi _{(\tau )}(\eta ),\psi _{(\tau )}(\xi )] \\
-\int_{\sigma _{x_{0}}^{(m-1)}}\bar{\Omega}_{(\tau )}^{(m-1),\intercal
}[\varphi _{(\tau )}(\eta ),\psi _{(\tau )}(\xi )]:=\Omega _{(x,\tau
)}^{\circledast }(\eta ,\xi )-\Omega _{(x_{0},\tau )}^{\circledast }(\eta
,\xi ),%
\end{array}%
\end{equation*}%
where, as before, $S(\sigma _{x}^{(m-1)},\sigma _{x_{0}}^{(m-1)})$ $\subset {%
\mathbb{R}}^{m}$ is a smooth hypersurface in the configuration space ${%
\mathbb{R}}^{m},$ spanned between two arbitrary but fixed nonintersecting
each other homological $(m-1)$-dimensional cycles $\sigma _{x}^{(m-1)}$and $%
\sigma _{x_{0}}^{(m-1)}\subset {\mathbb{R}}^{m},$ parametrized by points $x,$
$x_{0}\in {\mathbb{R}}^{m}.$ As a result of the construction above, the
Volterra type integral operators
\begin{equation}
\mathbf{\Omega }_{(\tau )}:=\mathbf{1}-\int_{\Sigma }d\rho (\xi
)\int_{\Sigma }d\rho (\eta )\tilde{\psi}_{(\tau )}(\xi )\Omega _{(x_{0},\tau
)}^{-1}(\xi ,\eta )\int_{{\mathcal{S}}_{-}(\sigma _{x}^{(m-2)},\sigma
_{x_{0}}^{(m-2)})}Z_{(\tau )}^{(m-1)}[\varphi _{(\tau )}(\eta ),\cdot ],
\label{4.13}
\end{equation}%
and
\begin{equation}
\mathbf{\Omega }_{(\tau )}^{\circledast }:=\mathbf{1}-\int_{\Sigma }d\rho
(\xi )\int_{\Sigma }d\rho (\eta )\tilde{\varphi}_{(\tau )}(\eta )\Omega
_{(x_{0},\tau )}^{\circledast ,-1}(\xi ,\eta )\int_{{\mathcal{S}}_{-}(\sigma
_{x}^{(m-2)},\sigma _{x_{0}}^{()m-2)})}\bar{Z}_{(\tau )}^{(m-1),\intercal
}[\cdot ,\psi _{(\tau )}(\xi )]  \label{4.14}
\end{equation}%
being bounded and invertible act, correspondingly, in the spaces ${\mathcal{H%
}}_{(\tau )}$ and ${\mathcal{H}}_{(\tau )}^{\ast }.$ Moreover, the Delsarte
transformed operator $\tilde{L}_{(\tau )}:{\mathcal{H}}_{(\tau
)}\longrightarrow {\mathcal{H}}_{(\tau )}$ can be written down as
\begin{equation}
\tilde{L}_{(\tau )}=\mathbf{\Omega }_{(\tau )}L_{(\tau )}\mathbf{\Omega }%
_{(\tau )}^{-1}=L_{(\tau )}+[\mathbf{\Omega }_{(\tau )},L_{(\tau )}]\mathbf{%
\Omega }_{(\tau )}^{-1},  \label{4.15}
\end{equation}%
being,\ due to reasoning as in \cite{SP,PSS}, also a differential
multi-dimensional operator in ${\mathcal{H}}_{(\tau )}.$

Now we can make the drawback reduction of our $\tau $-dependent objects,
recalling, that our operator (4.1) doesn't depend on the parameter $\tau \in
{\mathbb{R}}.$ In particular, from (\ref{4.8}) one can get that for any $%
(\varphi _{(\tau )}(\eta ),\psi _{(\tau )}(\xi ))\in {\mathcal{H}}_{(\tau
),0}^{\ast }\times {\mathcal{H}}_{(\tau ),0},$ $\xi ,\eta \in \Sigma ,$
\begin{equation}
\psi _{(\tau )}(\xi )=\psi _{\lambda }(\xi )e^{\lambda \tau },\qquad \varphi
_{(\tau )}(\eta )=\varphi _{\lambda }(\eta )e^{-\bar{\lambda}\tau }
\label{4.16}
\end{equation}%
with $\lambda \in \sigma (L)\cap \bar{\sigma}(L^{\ast })$ and any pair $%
(\varphi _{\lambda }(\xi ),\psi _{\lambda }(\xi ))\in {\mathcal{H}}%
_{0}^{\ast }\times {\mathcal{H}}_{0},$ $\xi ,\eta \in \Sigma _{\sigma },$
\begin{equation}
\begin{array}{c}
{\mathcal{H}}_{0}:=\{{\psi }_{\lambda }(\xi )\in {\mathcal{H}}%
_{-}:L(x;\partial |\lambda ){\psi }_{\lambda }(\xi )=0,\text{ \ } \\
{\psi }_{\lambda }(\xi )|_{\Gamma }=0,\text{ }(\lambda ;\xi )\in \sigma
(L)\cap \bar{\sigma}(L^{\ast })\mathbb{\times }\Sigma _{\sigma }\}, \\
{\mathcal{H}}_{0}^{\ast }:=\{\varphi _{\lambda }(\eta )\in {\mathcal{H}}%
_{-}^{\ast }:L^{\ast }(x;\partial |\lambda ){\varphi }_{\lambda }(\eta )=0,%
\text{ } \\
\varphi _{\lambda }(\eta )|_{\Gamma }=0,\text{ }(\lambda ;\eta )\in \sigma
(L)\cap \bar{\sigma}(L^{\ast })\mathbb{\times }\Sigma _{\sigma }\},%
\end{array}
\label{4.17}
\end{equation}%
where, by definition, $\Sigma _{\sigma }\times \mathbb{C}\subset \Sigma $ is
some "spectral"set of parameters. With respect to the closed subspaces ${%
\mathcal{H}}_{0}\subset {\mathcal{H}}_{-}$ and ${\mathcal{H}}_{0}^{\ast
}\subset {\mathcal{H}}_{-}^{\ast }$ the corresponding Delsarte transmutation
operators $\mathbf{\Omega }:{\mathcal{H}}\rightarrow {\mathcal{H}}$ and $%
\mathbf{\Omega }^{\circledast }:{\mathcal{H}}^{\ast }\rightarrow {\mathcal{H}%
}^{\ast }$ can be retrieved easily, making use of the expressions (\ref{4.15}%
), substituted into (\ref{4.12}) and (\ref{4.13}) :
\begin{equation}
\begin{array}{c}
\mathbf{\Omega }:=\mathbf{1}-\int_{\sigma (L)\cap \bar{\sigma}(L^{\ast
})}d\rho _{\sigma }(\lambda )\int_{\Sigma _{\sigma }}d\rho _{\Sigma _{\sigma
}}(\xi )\int_{\Sigma _{\sigma }}d\rho _{\Sigma _{\sigma }}(\eta )\tilde{\psi}%
_{\lambda }(\xi ) \\
\times \Omega _{x_{0}}^{-1}(\lambda ;\xi ,\eta )\int_{{\mathcal{S}}(\sigma
_{x}^{(m-1)},\sigma _{x_{0}}^{(m-1)})}Z^{(m)}[\varphi _{\lambda }(\eta
),\cdot ], \\
\mathbf{\Omega }^{\circledast }:=\mathbf{1}-\int_{\sigma (L)\cap \bar{\sigma}%
(L^{\ast })}d\rho _{\sigma }(\lambda )\int_{\Sigma _{\sigma }}d\rho _{\Sigma
_{\sigma }}(\xi )\int_{\sigma (L)\cap \bar{\sigma}(L^{\ast })}\int_{\Sigma
_{\sigma }}d\rho _{\Sigma _{\sigma }}(\eta )\tilde{\varphi}_{\lambda }(\eta )
\\
\times \Omega _{x_{0}}^{\circledast ,-1}(\lambda ;\xi ,\eta )\int_{{\mathcal{%
S}}(\sigma _{x}^{(m-1)},\sigma _{x_{0}}^{(m-1)})}Z^{(m)}[\cdot ,\psi
_{\lambda }(\xi )],%
\end{array}
\label{4.18}
\end{equation}%
where $d\rho _{\sigma }\times d\rho _{\Sigma _{\sigma }}$ is the
corresponding finite Borel measure on Borel subsets of $\sigma (L)\cap \bar{%
\sigma}(L^{\ast })\mathbb{\times }\Sigma _{\sigma }\subset \Sigma ,$
\begin{equation}
\begin{array}{c}
\tilde{\psi}_{\lambda }(\xi ):=\psi _{\lambda }(\xi )\cdot \Omega
_{x}^{-1}\Omega _{x_{0}} \\
\tilde{\varphi}_{\lambda }(\eta ):=\varphi _{\lambda }(\eta )\cdot \Omega
_{x}^{\circledast ,-1}\Omega _{x_{0}}^{\circledast },%
\end{array}
\label{4.19}
\end{equation}%
and due to semi-linearity, the expressions for kernels
\begin{equation}
\begin{array}{c}
\Omega _{x}(\lambda ;\xi ,\eta ):=\Omega _{(x,\tau )}[\varphi _{\lambda }e^{-%
\bar{\lambda}\tau },\psi _{\lambda }e^{\lambda \tau }], \\
\Omega _{x_{0}}[\varphi _{\lambda },\psi _{\lambda }]:=\Omega _{(x_{0},\tau
)}[\varphi _{\lambda }ee^{-\bar{\lambda}\tau },\psi _{\lambda }e^{\lambda
\tau }], \\
Z^{(m)}[\varphi _{\lambda },\psi _{\lambda }]:=Z_{(\tau )}^{(m)}[\varphi
_{\lambda }e^{-\bar{\lambda}\tau },\psi _{\lambda }e^{\lambda \tau }],%
\end{array}
\label{4.20}
\end{equation}%
don't depend on the whole on the parameter $\tau \in {\mathbb{R}}_{\tau }$
but only on $(\lambda ;\xi )\in \sigma (L)\cap \bar{\sigma}(L^{\ast })%
\mathbb{\times }\Sigma _{\sigma }.$ Moreover, if one to write down the
differential $m$-form $Z^{(m)}[\varphi _{\lambda },\psi _{\lambda }]\in $ $%
\Lambda ^{m}({\mathbb{R}}_{\tau }\times {\mathbb{R}}^{m};\mathbb{C}),$ $%
(\varphi _{\lambda }(\xi ),\psi _{\lambda }(\xi ))\in {\mathcal{H}}%
_{0}^{\ast }\times {\mathcal{H}}_{0},$ $\xi ,\eta \in \Sigma _{\sigma },$ as
\begin{eqnarray}
Z^{(m)}[\varphi _{\lambda },\psi _{\lambda }] &=&\sum_{i=1}^{m}dx_{1}\wedge
dx_{2}\wedge ...\wedge dx_{i-1}\wedge Z_{i}[\varphi _{\lambda },\psi
_{\lambda }]d\tau \wedge dx_{i+1}\wedge  \notag \\
&&...\wedge dx_{m}+Z_{0}[\varphi _{\lambda },\psi _{\lambda }]dx,
\label{4.21}
\end{eqnarray}%
then, due to the specially chosen $(m-1)$-dimensional homological cycles $%
(\sigma _{x}^{(m-1)}$ $,\sigma _{x_{0}}^{(m-1)})$ and the corresponding
closed m-dimensional surface $S(\sigma _{x}^{(m-1)},\sigma _{x_{0}}^{(m-1)})=%
{\mathbb{R}}^{m}$ at which $d\tau =0,$ the differential $m$-forms $%
Z^{(m)}[\varphi _{\lambda },\psi _{\lambda }]\in \Lambda ^{m}({\mathbb{R}}%
^{m};\mathbb{C}),$ $\lambda \in \sigma (L)\cap \bar{\sigma}(L^{\ast }),$
bring about the following expressions:
\begin{equation}
Z^{(m)}[\varphi _{\lambda },\psi _{\lambda }]=Z_{0}[\varphi _{\lambda },\psi
_{\lambda }]dx,  \label{4.22}
\end{equation}%
where, by definition, for any $i=\overline{0,m}$
\begin{equation}
Z_{i}[\varphi _{\lambda },\psi _{\lambda }]:=Z_{i,(\tau )}[\varphi _{\lambda
}e^{-\bar{\lambda}\tau },\psi _{\lambda }e^{\lambda \tau }],  \label{4.23}
\end{equation}%
being, evidently, not dependent more on the parameter $\tau \in {\mathbb{R}}$
but only on $\ \lambda \in \sigma (L)\cap \bar{\sigma}(L^{\ast }).$ Thus due
to (\ref{4.21}) and (\ref{4.22}) one can finally write down Delsarte
transmutation operators (\ref{4.12}) and (\ref{4.13}) as the following
invertible and bounded of Volterra type integral expressions:
\begin{equation}
\begin{array}{c}
\mathbf{\Omega }:=\mathbf{1}-\int_{\sigma (L)\cap \bar{\sigma}(L^{\ast
})}d\rho _{\sigma }(\lambda )\int_{\Sigma _{\sigma }}d\rho _{\Sigma _{\sigma
}}(\xi )\int_{\Sigma _{\sigma }}d\rho _{\Sigma _{\sigma }}(\eta )\tilde{\psi}%
_{\lambda }(\xi ) \\
\times \Omega _{x_{0}}^{-1}(\lambda ;\xi ,\eta )\int_{{\mathcal{S}}(\sigma
_{x}^{(m-1)},\sigma _{x_{0}}^{(m-1)})}Z_{(0)}^{(m)}[\varphi _{\lambda }(\eta
),\cdot ]dx, \\
\mathbf{\Omega }^{\circledast }:=\mathbf{1}-\int_{\sigma (L)\cap \bar{\sigma}%
(L^{\ast })}d\rho _{\sigma }(\lambda )\int_{\Sigma _{\sigma }}d\rho _{\Sigma
_{\sigma }}(\xi )\int_{\sigma (L)\cap \bar{\sigma}(L^{\ast })}\int_{\Sigma
_{\sigma }}d\rho _{\Sigma _{\sigma }}(\eta )\tilde{\varphi}_{\lambda }(\eta )
\\
\times \Omega _{x_{0}}^{\circledast ,-1}(\lambda ;\xi ,\eta )\int_{{\mathcal{%
S}}(\sigma _{x}^{(m-1)},\sigma _{x_{0}}^{(m-1)})}Z_{(0)}^{(m)}[\cdot ,\psi
_{\lambda }(\xi )]dx,%
\end{array}
\label{4.24}
\end{equation}%
where, by definition, $(\varphi _{\lambda },\psi _{\lambda })\in {\mathcal{H}%
}_{0}^{\ast }\times {\mathcal{H}}_{0},$ $\tilde{(\varphi }_{\lambda },\tilde{%
\psi}_{\lambda })\in \tilde{\mathcal{H}}_{0}^{\ast }\times \tilde{\mathcal{H}%
}_{0}$ and $\lambda \in \sigma (L)\cap \bar{\sigma}(L^{\ast }).$ The
operator expressions (\ref{4.24}) were defined before correspondingly, on
closed subspaces of generalized eigenfunctions ${\mathcal{H}}_{0}$ and ${%
\mathcal{H}}_{0}^{\ast }.$ In the case when these spaces are dense,
correspondingly, in ambient spaces $\mathcal{H}_{-}$ and $\mathcal{H}%
_{-}^{\ast }$ $,$ the operator expressions (\ref{4.24}) can be naturally
extended, correspondingly, upon the whole Hilbert spaces ${\mathcal{H}}_{-}$
and ${\mathcal{H}}_{-}^{\ast }$ as also invertible integral operators of \
Volterra type and, thereby, can be defined correspondingly on ${\mathcal{H}}
$ and ${\mathcal{H}}^{\ast }$ due to duality \cite{Be,BS} between Hilbert
spaces $\mathcal{H}_{-}$ and $\mathcal{H}_{+},$ where the latter space is
dense in $\mathcal{H}.$ The same way as in Chapter 3 above one can construct
the corresponding pair (\ref{3.4}) of Gelfand-Levitan-Marchenko integral
equations for an affine polynomial pencil (\ref{4.1}) of multi-dimensional
differential operators in the Hilbert space ${\mathcal{H}}.$ The latter note
finishes our present analysis of the structure of Delsarte transmutation
operators for pencils of multidimensional differential operators. Concerning
their natural applications to the inverse spectral problem and related
problems of feedback control theory mentioned before, we plan to stop on
them in more detail later.

\section{Acknowledgements}

Authors are cordially thankful to prof. B.N. Datta (USA, Illinois
University), prof. D.L. Blackmore (Newark USA, NJIT), prof. Nizhnik L.P.
(Kyiv, Inst. of Math.at NAS), prof. T. Winiarska (Krakow, PK), profs. A.
Pelczar and J. Ombach (Krakow, UJ) and prof. Z. Peradzynski (Warszawa, UW)
for valuable discussions of problems studied in the work.

\end{document}